\documentclass[shortnote,twocolumn]{jpsj3} %
\usepackage{txfonts}

\usepackage{bm}

\title{Cu NQR and NMR Studies of Optimally Doped Ca$_{2-x}$Na$_{x}$CuO$_{2}$Cl$_{2}$} 

\author{
Yutaka ITOH\thanks{E-mail address: yitoh@cc.kyoto-su.ac.jp}, 
Takato MACHI$^{1}$, 
Ikuya YAMADA$^{2}$,
Masaki AZUMA$^{3}$, 
and Mikio TAKANO$^{4}$
}
\inst{\address{Department of Physics, Graduate School of Science, Kyoto Sangyo 
University, Kamigamo-Motoyama, Kika-ku Kyoto 603-8555, Japan} \\
{$^{1}$Superconductivity Research Laboratory, International Superconductivity Technology Center, 1-10-13 Shinonome, Koto-ku Tokyo 135-0062, Japan}\\
{$^{2}$Osaka Prefecture University, Sakai, Osaka 599-8531, Japan}\\
{$^{3}$Materials and Structures Laboratory, Tokyo Institute of Technology, Nagatsuta, Midori-ku, Yokohama 226-8503, Japan}\\
{$^{4}$Institute for Chemical Research, Kyoto University, Uji, Kyoto 611-0011, Japan}\\
}

\date{\today}

\kword{Ca$_{2-x}$Na$_{x}$CuO$_{2}$Cl$_{2}$, nuclear quadrupole resonance, NMR, superconductor}

\begin{document}
\maketitle
 
Copper oxychloride Ca$_{2-x}$Na$_{x}$CuO$_{2}$Cl$_{2}$ 
is a single-CuO$_2$-layer system of high-$T_\mathrm{c}$ cuprate superconductors \cite{Hiroi1,Hiroi2}.
The parent compound Ca$_{2}$CuO$_{2}$Cl$_{2}$ is one of the ideal square lattice antiferromagnets
with no orthorhombic distortion.
Apical chlorine distant from the CuO$_2$ plane is characteristic of this system.  
For Na substitution, the crystal symmetry is still tetragonal, and 
the oxygen and chlorine composition is robust. 
The flat CuO$_2$ plane is similar to HgBa$_2$CuO$_{4+\delta}$.   
The out-of-plane disorder of the Na dopant for the Ca site primarily causes randomness effect. 
A higher $T_\mathrm{c}$ has been associated with the flatness of CuO$_2$ planes in a unit cell. 
However,  the optimized $T_\mathrm{c}$ $\approx$ 28 K with $x\approx$ 0.2 is lower than the highest $T_\mathrm{c}$ $\approx$ 98 K of HgBa$_2$CuO$_{4+\delta}$.  
The single crystals grown at high pressure \cite{Azuma}
were served for scanning tunneling spectroscopy (STS) studies, which revealed exotic electronic states \cite{Hanaguri, Kohsaka, Hanaguri2}. 
 
In this note, we report on $^{63, 65}$Cu nuclear quadrupole resonance (NQR) and NMR studies of an optimally doped superconductor Ca$_{2-x}$Na$_{x}$CuO$_{2}$Cl$_{2}$ ($T_\mathrm{c}$ $\approx$ 28 K for $x\approx$ 0.2).
Although the system has a robust oxygen composition, 
we observed a multiple Cu NQR frequency spectrum and broad Cu NMR spectra. 
We also observed nonexponential nuclear spin-lattice relaxation. 
Some inhomogeneous features are similar to those for La$_{2-x}$Sr$_{x}$CuO$_{4-\delta}$.

A polycrystalline sample of Ca$_{1.8}$Na$_{0.2}$CuO$_{2}$Cl$_{2}$ was synthesized under high pressure, as described in refs. 1 and 2. 
The sample was confirmed to be in the single phase from its powder X-ray diffraction pattern. 
For the NMR and NQR experiments, the grains of the powder were aligned in a magnetic field of about 8 T and were fixed in an epoxy resin (Staycast 1266) \cite{Takigawa}.  
 
A phase-coherent-type pulsed spectrometer was utilized to perform the $^{63, 65}$Cu 
NQR and NMR (nuclear spin $I$ = 3/2) experiments for the $c$-axis-aligned sample.    
The spectra were obtained by recording the spin-echo intensity against frequency point by point. 
Cu nuclear spin-lattice relaxation curves $p(t)\equiv 1-M(t)/M(\infty)$ (recovery
curves) were measured by an inversion recovery technique as a function of time $t$ after an inversion pulse,  where the nuclear
spin-echo $M(t)$, $M(\infty)[\equiv M(10T_1)]$, and $t$ were recorded. 

The recovery curves in the Cu NQR were analyzed by an exponential function with $T_{1}$ multiplied by  a stretched exponential function with $\tau_{1}$,  $p(t)=p(0){e}^{-3t/T_1-\sqrt{3t/\tau_1}} $, as in refs. 8 and 9. 
The finite 1/$\tau_{1}$ indicates the existence of random relaxation centers, e.g., local moments of dilute magnetic impurities, and spinless defects in strongly correlated systems.  $T_{1}$ is due to the host electron spin relaxation process. 
 
\begin{figure}[h]
 \begin{center}
 \includegraphics[width=0.8\linewidth]{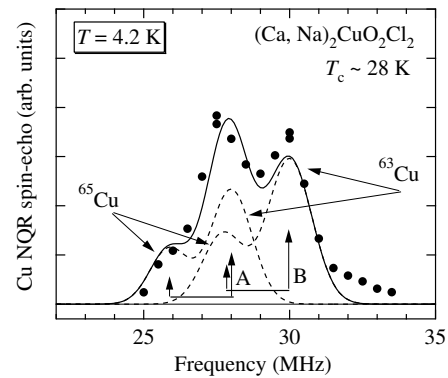}
 \end{center}
 \caption{\label{fig:NQR}
Zero-field Cu NQR spectrum (solid circles) of Ca$_{2-x}$Na$_{x}$CuO$_{2}$Cl$_{2}$ ($T_\mathrm{c}$ $\approx$ 28 K, $x \approx$ 0.2) at $T$ = 4.2 K. Dashed curves are guides for the eye showing double Gaussian functions
for a pair of $^{63, 65}$Cu isotopes.  
A solid curve is a simulation with the $^{63}$Cu peaks at $f_\mathrm{A}$ = 28.0 MHz and $f_\mathrm{B}$ = 30.0 MHz. 
 }
 \end{figure}

Figure \ref{fig:NQR} shows the zero field $^{63, 65}$Cu NQR frequency spectrum at $T$ = 4.2 K.
Solid and dashed curves are guides for the eye showing Gaussian functions.
A pair of $^{63, 65}$Cu isotope NQR signals is assigned to a Cu site.   
The natural abundance ratio of the two $^{63, 65}$Cu isotopes is 0.69:0.31.
The quadrupole moment ratio of the two $^{63, 65}$Cu isotopes is $^{63}Q$/$^{65}Q$ = 1.08.
A pair of Gaussian functions with the relevant intensity and resonance frequency ratios is assumed for a Cu site.
In Fig. \ref{fig:NQR},  two Cu sites are assumed to reproduce the observed NQR spectrum,  
although the decomposition is not unique.  
 
The multiple features are similar to those for La$_{2-x}$Sr$_x$CuO$_{4-\delta}$ \cite{Yoshimura,Yoshimura1,Imai1,Imai2} and La$_{2}$CuO$_{4+\delta}$ \cite{Hammel}.
Randomness in crystalline potentials is not sufficient to cause such a multiple NQR spectrum \cite{ItohYSCO, NQR0}. 
Some modulation in the charge density of the CuO$_2$ planes may be induced by the doped holes.  
The symmetry of the charge density may be broken to become lower than the tetragonal symmetry of the original square lattice. 
The electronic cluster glass in the STS observation \cite{Kohsaka} is such a misfit state, and can cause the multiple broad Cu NQR spectrum.   
    
\begin{figure}[h]
 \begin{center}
 \includegraphics[width=1.0\linewidth]{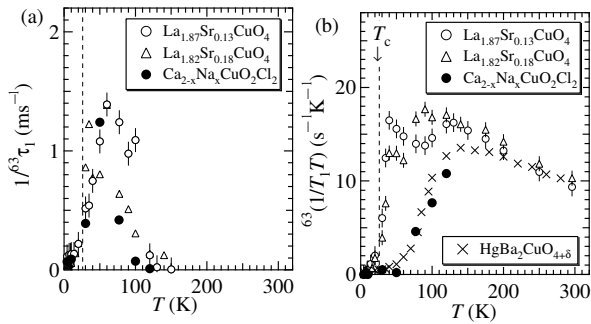}
 \end{center}
 \caption{\label{fig:1/T1}
Temperature dependences of Cu nuclear spin-lattice relaxation rates 
1/$\tau_{1}$ (a) and 1/$T_{1}T$ (b) at $f_\mathrm{A}$ = 28.0 MHz for Ca$_{1.8}$Na$_{0.2}$CuO$_{2}$Cl$_{2}$.
1/$\tau_{1}$ and 1/$T_{1}T$ (open circles and open triangles) of $^{63}$Cu(A) of La$_{2-x}$Sr$_x$CuO$_{4-\delta}$ ($x$ = 0.13 and 0.18) are reproduced from ref. 9. 
1/$T_{1}T$ of $^{63}$Cu of optimally doped HgBa$_{2}$CuO$_{4+\delta}$ ($T_\mathrm{c}$ = 96 K) 
is reproduced from refs. 17 and 18. 
 }
 
 \end{figure}

Figure \ref{fig:1/T1} shows temperature dependences of Cu nuclear spin-lattice relaxation rates
of the Cu NQR ($f_\mathrm{A}$ = 28.0 MHz) for Ca$_{1.8}$Na$_{0.2}$CuO$_{2}$Cl$_{2}$,
(a) the stretched exponential relaxation rate 1/$\tau_{1}$ and
(b) the exponential relaxation rate divided by $T$, 1/$T_{1}T$. 
For comparison, 1/$\tau_{1}$ and 1/$T_{1}T$ for La$_{2-x}$Sr$_x$CuO$_{4-\delta}$ ($x$ = 0.13 and 0.18) \cite{Itoh2} and for HgBa$_{2}$CuO$_{4+\delta}$ ($T_\mathrm{c}$ = 96 K)  \cite{Itoh3,Itoh4} are also shown.
 \\
 \\
 
In Fig.  \ref{fig:1/T1}(a), 1/$\tau_{1}$ of  Ca$_{1.8}$Na$_{0.2}$CuO$_{2}$Cl$_{2}$ is similar to those of La$_{2-x}$Sr$_x$CuO$_{4-\delta}$.  The appreciable 1/$\tau_{1}$ indicates the existence of a slow fluctuation mode in a low-frequency spin fluctuation spectrum, which could cause the pair-breaking effect \cite{Itoh,Itoh2}.
The electronic cluster glass state \cite{Kohsaka} can be the low-frequency local modes \cite{Itoh2}.   
 
 \begin{figure}[h]
 \begin{center}
 \includegraphics[width=1.0\linewidth]{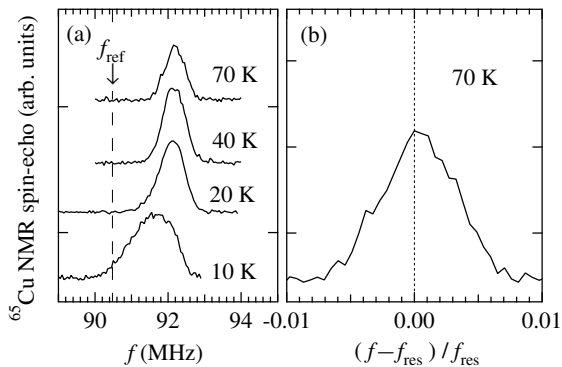}
 \end{center}
 \caption{\label{fig:NMR}
(a) Temperature dependence of central transition lines ($I_z$ = 1/2 $\leftrightarrows$ -1/2) of $^{65}$Cu NMR frequency spectra for Ca$_{1.8}$Na$_{0.2}$CuO$_{2}$Cl$_{2}$
in a magnetic field of 7.48414 T along the $c$ axis.
$f_\mathrm{ref}$ = 90.4758 MHz denotes the zero shift resonance frequency. 
(b) $^{65}$Cu NMR spectrum at 70 K against the normalized frequency shift defined by
$(f-f_\mathrm{res})/f_\mathrm{res}$ with $f_\mathrm{res}$ = 92.15 MHz.  
 }
 \end{figure}  
 
In contrast to the random relaxation process of 1/$\tau_{1}$, the uniform relaxation process of 1/$T_{1}T$ for Ca$_{1.8}$Na$_{0.2}$CuO$_{2}$Cl$_{2}$ is different from those of La$_{2-x}$Sr$_x$CuO$_{4-\delta}$, but similar to that of the optimally doped HgBa$_{2}$CuO$_{4+\delta}$ ($T_\mathrm{c}$ = 96 K).  

Figure \ref{fig:NMR}(a) shows the central transition lines ($I_z$ = 1/2 $\leftrightarrows$ -1/2) of $^{65}$Cu NMR frequency spectra for Ca$_{1.8}$Na$_{0.2}$CuO$_{2}$Cl$_{2}$
in a magnetic field of 7.48414 T applied along the $c$ axis. 
We focused on $^{65}$Cu NMR experiments, because
the nuclear gyromagnetic ratio $^{63}\gamma_\mathrm{n}$ of $^{63}$Cu is close to that of $^{23}$Na,
making their NMR signals indistinguishable.
The $^{65}$Cu NMR spectrum was nearly independent of temperature. 
The Knight shift $^{65}K_{cc}$ was estimated to be $\sim$1.85 $\%$ in the temperature range of 20$-$100 K.  
 
The NMR linewidth at $T$ = 10 K is broader than those above 20 K.  
This is due to the distribution of local internal fields of a vortex lattice in the mixed state
of the type II superconductor. The vortex pattern is unclear in the present NMR line profile;
hence it might be due to a Bragg glass. 
 
Figure \ref{fig:NMR}(b) shows the $^{65}$Cu NMR spectrum at  70 K against the normalized frequency shift defined by $(f-f_\mathrm{res})/f_\mathrm{res}$ with the peak frequency $f_\mathrm{res}$.   
The broad linewidth of Ca$_{1.8}$Na$_{0.2}$CuO$_{2}$Cl$_{2}$ is about 5 times wider than that of HgBa$_2$CuO$_{4+\delta}$ \cite{Itoh4} and
is also observed for La$_{2-x}$Sr$_x$CuO$_{4-\delta}$  \cite{ItohLSCO}. 
Thus, a common electronic inhomogeneity is observed in Ca$_{1.8}$Na$_{0.2}$CuO$_{2}$Cl$_{2}$ and La$_{2-x}$Sr$_x$CuO$_{4-\delta}$ \cite{Itoh2,Imai1,Imai2,ItohLSCO}.
 
 \acknowledgments  
We thank K. Oka for experimental support. 
Y. I. thanks H. Kageyama for sample preparation at an early stage. 
This study was supported in part by Grant-in-Aid for Scientific Research (C), from the Ministry of 
Education, Culture, Sports, Science and Technology of Japan (Grant No. 22540353).

\end{document}